# Three Sides of Retrieval: Factorial Evidence for Document-Side, Query-Side, and Answer-Side Complementarity in RAG

Ng S. T. Chong
Campus Computing Centre, United Nations University
Tokyo, Japan
ngstc@unu.edu

**Abstract:** RAG systems rely on chunking, which destroys structural information in documents. Existing heading-based retrieval (Jeong et al., 2025) requires multiple LLM calls per document and returns sub-chunks within matched sections. We introduce ToC-guided page retrieval, which infers headings from visual formatting without LLM calls, embeds them as a parallel index, and loads full page sections. Across 1,280 conditions on 8 enterprise documents (5 to 195 pages), we find: (1) ToC is a significant main effect on answer quality ($d = +0.41$, $p = 0.031$), with the largest gains in completeness (+0.40) and usefulness (+0.40); (2) combined with answer-side verification, it outperforms query-side decomposition + verification ($d = +0.32$, $p = 0.036$); (3) ToC contributes 20% of citations despite adding only 2.9 pages per query; (4) gains are directionally larger on longer documents (up to +1.50 on 118 pages), though the trend does not reach significance with 8 documents; and (5) a 480-condition sensitivity analysis finds no significant parameter effects (all $p > 0.38$), confirming defaults are near-optimal. The contribution is both methodological (a new zero-LLM-cost retrieval algorithm) and empirical: factorial evidence that document-side, query-side, and answer-side enhancements are complementary, a three-way interaction not previously studied.

## 1. Introduction

Chunking is the universal first step in RAG: documents are split into fixed-size segments (typically 256–512 tokens) for vector embedding and retrieval. While effective for matching individual passages, chunking systematically destroys the structural information encoded in document organization, including heading hierarchies, section boundaries, and cross-references between related content.

The most advanced RAG techniques address problems at two of three stages. On the *query side*, Agentic Search (AS) decomposes complex questions into sub-queries for parallel retrieval (Trivedi et al., 2023; Ma et al., 2023). On the *answer side*, Completeness Check (CC) verifies coverage and triggers re-retrieval for gaps (Asai et al., 2024; Shao et al., 2023). Neither addresses the *document side*: when a query requires content from multiple related sections, chunk-based retrieval may miss structurally relevant passages that are semantically distant from the query but organizationally proximate in the source document.

The idea of using document headings for retrieval is not new. DocsRay (Jeong et al., 2025) generates pseudo tables of contents via LLM and uses them for section-level retrieval. However, DocsRay's LLM-dependent heading generation adds cost and introduces heading quality as a confound, and its within-section chunk retrieval discards the cross-paragraph context that section-level matching is positioned to provide. We implement Table-of-Contents-guided page retrieval (ToC), which takes a different approach at each stage: it extracts headings without LLM calls, using native PDF bookmarks when available or scanning every page through a layout analysis engine that infers heading hierarchy from visual formatting (font sizes, weights, spacing) when no bookmarks exist, and loads full page sections rather than drilling into sub-chunks. The layout analysis approach is not a crude fallback; it examines typographic signals across the entire document to reconstruct a hierarchical table of contents from visual cues, succeeding on all eight documents in our corpus including those without PDF bookmark metadata. Combined with full-page loading and a parallel index architecture, this produces a system that is both cheaper and more context-preserving than LLM-generated heading approaches, while enabling clean factorial measurement of heading-based retrieval in isolation.

We evaluate ToC through four experiments on a production enterprise RAG system serving HR, procurement, and travel-policy documents:

- **Experiment 1** ($2^3$ factorial, 384 conditions): Quantifies ToC's main effect and interactions with Section Expansion (SE) and Completeness Check (CC).
- **Experiment 2** (head-to-head, 384 conditions): Compares ToC against Agentic Search (AS) and CC in pairwise and combined configurations.
- **Experiment 3** (document-length scaling, 32 conditions): Tests whether ToC's benefit scales with document complexity.
- **Experiment 4** (parameter sensitivity, 480 conditions): Sweeps 60 parameter combinations to test robustness of default settings.

Our contribution is both methodological and empirical. Methodologically, we introduce ToC-guided page retrieval: a document-side retrieval algorithm that extracts headings from document structure without LLM calls, embeds them as a parallel retrieval index, matches queries against headings via cosine similarity, and loads full page sections for matched headings. This differs from prior heading-based approaches in extraction method (layout-inferred vs LLM-generated), retrieval granularity (full pages vs within-section chunks), and index architecture (parallel channels vs score blending). Empirically, we provide factorial evidence that document-side retrieval (ToC) and answer-side verification (CC) are complementary and additive (no significant interaction, $p = 0.317$), that their combination outperforms the query-side + answer-side pairing (AS+CC) by $d = +0.32$, and that full-page loading improves answer quality despite the coarser retrieval granularity, a non-obvious result since adding broader context could plausibly introduce noise and reduce accuracy.

## 2. Related Work

### *2.1 Chunking and Its Limitations*

Lewis et al. (2020) introduced RAG with fixed-size chunking as the default retrieval unit. Barnett et al. (2024) catalog seven failure points in RAG pipelines, including missing content and incomplete answers that trace to chunking fragmentation. Liu et al. (2024b) demonstrate that position within retrieved context affects utilization ("lost in the middle"), further compounding the problem when relevant content is split across chunks.

### *2.2 Query-Side Enhancements*

Query decomposition addresses retrieval by reformulating the user's question. IRCoT (Trivedi et al., 2023) interleaves retrieval with chain-of-thought reasoning across multiple hops. Ma et al. (2023) rewrite queries for better alignment with passage embeddings. Gao et al. (2024) survey modular RAG architectures categorizing features into pre-retrieval, retrieval, and post-retrieval stages. These approaches improve what is searched for but not what is available to be found, since they operate on the same chunk index.

### *2.3 Answer-Side Verification*

Self-RAG (Asai et al., 2024) uses reflection tokens to decide when re-retrieval is needed. ITER-RETGEN (Shao et al., 2023) iterates retrieval-generation cycles. These post-generation methods catch gaps that retrieval missed but incur additional LLM calls per verification cycle, adding latency and cost.

### *2.4 Document-Structure-Aware Retrieval*

Several recent approaches exploit document structure for retrieval. RAPTOR (Sarthi et al., 2024) builds a tree through recursive clustering and LLM summarization, a bottom-up approach that creates structure at substantial LLM cost during indexing. TreeRAG (Tao et al., 2025) constructs LLM-generated hierarchical trees with subtitles and index markers, requiring LLM calls at both indexing and retrieval.

DocsRay (Jeong et al., 2025) is the closest prior work. DocsRay generates a pseudo table of contents through a multi-phase LLM process: the document is divided into 5-page chunks, an LLM performs binary boundary detection between each consecutive pair, small sections are merged based on embedding similarity, and the LLM generates a title for each resulting section. Section titles and content embeddings are combined using weighted interpolation (a blending parameter β) for chunk retrieval within matched sections. DocsRay was evaluated on documents averaging 49.4 pages. Our approach shares the core insight that heading-level matching serves as a retrieval signal, but advances the design in three dimensions:

1. **Heading extraction: layout-inferred vs. LLM-generated.** DocsRay requires multiple LLM calls per document: one boundary detection call per 5-page chunk pair plus one title generation call per section. For a 195-page document, this means approximately 40 LLM calls at indexing time. This approach works on any document but adds substantial cost and introduces heading quality as a confounding variable. We extract headings without LLM calls, using a three-tier fallback: (a) native PDF bookmarks/outlines when present, (b) page-by-page layout analysis that converts each page to structured markdown by examining font sizes, weights, and spacing to infer heading hierarchy, effectively reconstructing a table of contents from visual formatting cues across the entire document, or (c) first-line extraction as a last resort. The layout analysis tier is not a simple pattern-matching heuristic: it scans every page for typographic signals that distinguish headings from body text, producing a hierarchical heading structure even when the PDF contains no bookmark metadata. This succeeds on all eight documents in our corpus, including those without PDF bookmarks, and would extend to any document with visual heading formatting (the vast majority of enterprise, institutional, and published documents). It would fail only on completely unformatted text (e.g., plain-text exports, chat logs). The result is a heading extraction method that covers a broad class of real-world documents at zero LLM cost, while eliminating LLM heading quality as a variable in our experiments.

2. **Retrieval granularity: full pages vs. within-section chunks.** DocsRay matches headings but then retrieves individual chunks within matched sections, preserving section matching at the cost of re-fragmenting the very content that section-level retrieval was meant to keep intact. We load the entire page section, preserving cross-paragraph context: definitions, conditions, and exceptions that appear together in the source document. This coarser granularity could plausibly hurt accuracy by including irrelevant paragraphs, but our data shows the opposite: completeness improves (+0.20) with no accuracy degradation (+0.11). Full-page loading is not just a simpler design; it delivers the context-preservation benefit that motivates section-level retrieval in the first place.
3. **Index architecture: parallel channels vs. score blending.** DocsRay interpolates heading and content similarity scores using a tunable blending parameter β, coupling the two signals and requiring parameter optimization. Our headings and chunks operate as independent retrieval channels, each contributing its own evidence to the context without score manipulation. This architecture is simpler to deploy (no β to tune), more transparent (each channel's contribution is directly measurable), and empirically effective: 20% of citations come from ToC-loaded pages, confirming the channel adds unique evidence that chunk retrieval alone misses.

BookRAG (Wang et al., 2025) builds a hierarchical tree index from document structure and entity graphs, using an LLM agent to traverse at query time. KohakuRAG (Yeh et al., 2026) implements a four-level tree with LLM-powered query planning. HiSem-RAG (2026) preserves hierarchical semantic relationships in chunking. DeepRead (Li et al., 2026) employs structure-aware agentic search with locate-then-read reasoning. All require LLM involvement at indexing or query time, or both.

Dense Hierarchical Retrieval (Liu et al., 2021) pioneered two-stage document-then-passage ranking but requires separate document-level training. LlamaIndex's DocumentSummaryIndex embeds LLM-generated summaries per document for coarse retrieval, similar in concept but at the document level with LLM cost.

Our contribution builds on the insight, established by DocsRay, that headings are useful for retrieval, but advances it in two directions: (1) an alternative design that eliminates LLM cost through layout-inferred extraction and preserves cross-paragraph context through full-page loading, with empirical evidence that the coarser granularity improves rather than degrades answer quality; and (2) factorial empirical evidence of how heading-based retrieval interacts with query-side decomposition and answer-side verification, a three-way interaction not previously studied despite growing interest in each technique individually.

## 3. Two Blind Spots in Chunk-Based Retrieval

We identify two structural retrieval failures that motivate ToC-guided retrieval. Both arise from chunking fragmentation and cannot be addressed by query-side or answer-side techniques, since those operate on the same chunk index.

### *3.1 Cousin-Section Gap*

Section Expansion (SE), which retrieves sibling chunks that share the same heading breadcrumb, is a common approach to recovering local context. However, SE is limited to chunks within the same heading sub-tree. When a query requires content from *cousin sections* (sibling sub-trees under a common parent), SE cannot bridge the gap. For example, a question about "all types of leave" may require content from sections on annual leave, sick leave, and home leave, each under separate headings that share a parent heading ("Leave") but whose chunks are not siblings.

### *3.2 Coarse-Heading Gap*

Documents with flat section structures (long sections under a single heading with no sub-headings) fragment into chunks that are semantically distant from each other despite being part of the same topical unit. A query about the section's overall topic may retrieve only one or two chunks from the middle, missing the introductory definitions or concluding conditions that appear in other chunks from the same section. The heading itself captures the section's topic at a level of abstraction that individual chunks cannot.

### *3.3 Why Query-Side and Answer-Side Approaches Cannot Fix These*

Agentic Search decomposes queries into sub-queries, but each sub-query still retrieves from the same chunk index. If the relevant chunks are semantically distant from all reasonable sub-queries, decomposition cannot find them. Completeness Check detects gaps in the answer and triggers re-retrieval, but re-retrieval also searches the same chunk index and may find the same (incomplete) set of chunks repeatedly.

ToC addresses these gaps by providing an alternative retrieval path: matching the query against section-level headings and loading the full page content of matched sections, bypassing the chunk layer entirely.

## 4. ToC-Guided Page Retrieval

### *4.1 Architecture*

**Ingestion.** During document processing, headings are extracted using a three-tier fallback: (1) native PDF bookmarks/outlines if present, (2) page-by-page layout analysis that converts each page to structured markdown by examining font sizes, weights, and spacing to infer heading hierarchy, scanning the entire document to reconstruct a table of contents from visual cues, or (3) first-line extraction as a last resort for documents with no detectable heading structure. No LLM calls are used. Each extracted heading is embedded using the same embedding model as chunk embeddings. The heading text, its embedding vector, and the page numbers it spans are stored in a heading index (~10–50 entries per document). Separately, full page text is stored for direct loading.

**Query time.** The user's query embedding is compared against all heading embeddings via cosine similarity. Headings exceeding a minimum similarity threshold (default 0.3) are ranked, and the top-k matches (default 3) are selected. The full page text for matched sections is loaded and prepended to the chunk-retrieved context, up to a page budget (default 5 pages). Deduplication removes any overlap between ToC-loaded pages and chunk-retrieved content.

### *4.2 Design Decisions*

Beyond layout-inferred heading extraction, two further design choices distinguish ToC from prior heading-based approaches:

**Full-page loading, not sub-page filtering.** When a heading matches, the entire page section is loaded rather than drilling into sub-chunks. Prior work (DocsRay) matches headings but then retrieves individual chunks within matched sections, re-fragmenting the content that section-level matching was meant to preserve. Full-page loading preserves cross-paragraph context: definitions, conditions, and examples that appear together in the source document. The trade-off is coarser granularity: irrelevant paragraphs within a matched section are included. Whether the additional context helps or dilutes is an empirical question we address in Section 6.3; the answer is unambiguously positive.

**Parallel index, not score blending.** The heading index operates alongside the chunk index as an independent retrieval channel. Each contributes separately to the final context: chunks provide precise passage-level matches, while ToC provides section-level structural context. This avoids the score interpolation and parameter tuning required by blended approaches (DocsRay's β), and makes it possible to measure each channel's contribution directly (Section 6.3).

### *4.3 Cost Profile*

Table 1 compares the cost profile of ToC against query-side (AS) and answer-side (CC) approaches, as well as DocsRay's LLM-generated heading approach.

**Table 1.** Cost comparison of RAG enhancement techniques. DocsRay costs are from Jeong et al. (2025).

| Metric | ToC (ours) | DocsRay | AS | CC |
|---|---|---|---|---|
| LLM calls at indexing | 0 | ~P/5 boundary + S title per doc | 0 | 0 |
| LLM calls at query time | 0 | 0 | 1 (decomposition) + N (sub-queries) | 1–2 per verification |
| Embedding API calls at query time | 0 (pre-embedded) | 0 (pre-embedded) | N (sub-query embeddings) | 1+ (re-retrieval) |
| Parameter tuning required | None | β blending weight | None | None |
| Typical latency added | <1s | <1s | 5–15s | 30–120s |
| Measured avg latency (Exp 2) | 60s total | — | 68s total | 177s total |

P = page count, S = number of detected sections. For a 195-page document, DocsRay requires approximately 40 LLM calls at indexing time (38 boundary detections + title generation for each section). ToC's only runtime cost is in-memory cosine similarity against ~10–50 heading vectors per document, which is negligible, and it incurs zero LLM cost at indexing time.

## 5. Experiments

### *5.1 System and Corpus*

We evaluate on a production enterprise RAG system serving an internal knowledge base across four domains: HR, procurement, ERP, and travel policy. The corpus comprises 8 documents ranging from 5 to 195 pages. Three are publicly available: the UN Staff Regulations and Rules (118 pages, HR), the UNOPS Procurement Manual (181 pages, ERP), and the EU General Data Protection Regulation (78 pages, IT). The remaining five are internal institutional guides and policies (5–48 pages each, covering HR, ERP, and travel). All documents have detectable heading structure (PDF bookmarks or formatted headings) suitable for rule-based ToC extraction.

The retrieval pipeline uses fixed-size chunking (512 tokens, 64-token overlap) with breadcrumb-enriched chunks, embedded using nomic-embed-text (768 dimensions). Retrieval combines cosine similarity and BM25 keyword matching via PostgreSQL/pgvector, followed by cross-encoder reranking. The system chains fixed stages (query safety screening, domain classification, hybrid retrieval, reranking, and LLM generation) with four optional enhancement features: Section Expansion (SE), Agentic Search (AS), Completeness Check (CC), and Table-of-Contents-guided retrieval (ToC).

### *5.2 Evaluation Method*

All answers are scored using model-answer anchored evaluation. For each query, a reference answer is generated using the RAG system with all features enabled, then manually verified for factual correctness against the source documents. A separate evaluator model (Claude Opus 4.6) scores candidate answers against these verified references on three dimensions (1–5 Likert scale): accuracy (factual correctness vs. reference), completeness (coverage of reference content), and usefulness (actionability for a practitioner). Composite quality is the unweighted mean of the three dimensions.

Two answer-generating models are used in Experiments 1-3: DeepSeek-V4-Flash and grok-4.3, both via Azure AI. Experiment 4 uses DeepSeek-V4-Flash only. Statistical tests use paired Wilcoxon signed-rank tests with FDR correction; effect sizes are Cohen's d with 10,000-iteration bootstrap 95% confidence intervals.

### *5.3 Experiment 1: $2^3$ Factorial (SE × ToC × CC)*

**Design.** A 2×2×2 full factorial crossing SE, ToC, and CC (AS excluded to isolate ToC without the embedding-consumption confound described in Section 3.3). Table 2 shows the 8 configurations; each is evaluated on 24 queries × 2 models = 384 conditions.

**Queries.** 24 queries stratified across 4 types: simple single-facet (n=6), multi-section within one document (n=6), cousin-section gap (n=6), and coarse-heading gap (n=6).

**Table 2.** Experiment 1 configuration matrix.

| # | Config | SE | ToC | CC |
|---|---|---|---|---|
| 0 | Baseline | - | - | - |
| 1 | SE | ON | - | - |
| 2 | ToC | - | ON | - |
| 3 | CC | - | - | ON |
| 4 | SE+ToC | ON | ON | - |
| 5 | SE+CC | ON | - | ON |
| 6 | ToC+CC | - | ON | ON |
| 7 | SE+ToC+CC | ON | ON | ON |

### *5.4 Experiment 2: Head-to-Head (ToC vs AS vs CC)*

**Design.** Compares ToC against AS and CC in isolation and in combinations, including the key comparison of document-side + answer-side (ToC+CC) vs. query-side + answer-side (AS+CC). Table 3 shows the 8 configurations; each is evaluated on 24 queries × 2 models = 384 conditions.

**Table 3.** Experiment 2 configuration matrix.

| # | Config | ToC | AS | CC |
|---|---|---|---|---|
| 0 | Baseline | - | - | - |
| 1 | AS | - | ON | - |
| 2 | CC | - | - | ON |
| 3 | ToC | ON | - | - |
| 4 | AS+CC | - | ON | ON |
| 5 | ToC+CC | ON | - | ON |
| 6 | ToC+AS | ON | ON | - |
| 7 | ToC+AS+CC | ON | ON | ON |

### *5.5 Experiment 3: Document-Length Scaling*

**Design.** Baseline vs. ToC on queries targeting individual documents across the full page-count range (5–195 pages). 2 configurations × 8 queries × 2 models = 32 conditions.

### *5.6 Experiment 4: Parameter Sensitivity*

**Design.** A grid sweep of the three ToC parameters: page budget (1, 3, 5, 7, 10), minimum similarity threshold (0.15, 0.25, 0.35, 0.45), and top-k headings (1, 3, 5). All 60 parameter combinations are evaluated on 8 representative queries (2 per query type) using a single model (DeepSeek-V4-Flash), yielding 480 conditions. Appendix B.3 lists the full parameter grid.

## 6. Results

### *6.1 Experiment 1: ToC Is a Significant Main Effect*

Three-way ANOVA on composite quality reveals ToC as a significant main effect ($F = 4.69$, $p = 0.031$, $\eta^2 p = 0.012$). CC is the strongest main effect ($F = 12.03$, $p < 0.001$, $\eta^2 p = 0.031$). SE is not significant ($F = 0.09$, $p = 0.762$). No two-way or three-way interactions reach significance, indicating the features are additive rather than competing. Both effect sizes are small by Cohen's benchmarks (small = 0.01, medium = 0.06), reflecting that most variance in composite quality is driven by the query and document, since some queries are inherently harder regardless of configuration. The practical implication is that ToC and CC are independently useful levers whose contributions stack: CC contributes roughly 2.5× more explained variance than ToC, but the best configuration simply turns both on.

Table 4 reports quality scores for all 8 configurations. ToC+CC achieves the highest composite quality (4.22), outperforming all other configurations including the full feature set SE+ToC+CC (4.14). Figure 1 visualizes these results with 95% confidence intervals.

**Table 4.** Experiment 1: Quality by configuration (accuracy, completeness, usefulness, composite, and latency; 1–5 scale, averaged across both models).

| Config | Accuracy | Completeness | Usefulness | Composite | Latency |
|---|---|---|---|---|---|
| Baseline | 3.65 | 3.50 | 4.00 | 3.72 | 66s |
| SE | 3.75 | 3.58 | 4.08 | 3.81 | 76s |

| ToC | 3.83 | 3.90 | 4.40 | 4.04 | 73s |
|---|---|---|---|---|---|
| CC | 3.83 | 3.96 | 4.50 | 4.10 | 165s |
| SE+ToC | 3.77 | 3.79 | 4.29 | 3.95 | 66s |
| SE+CC | 3.92 | 3.92 | 4.44 | 4.09 | 146s |
| ToC+CC | 4.00 | 4.10 | 4.56 | 4.22 | 171s |
| SE+ToC+CC | 3.98 | 3.96 | 4.48 | 4.14 | 154s |

Notable patterns in Table 4: ToC's largest contribution is to completeness (+0.40 over baseline) and usefulness (+0.40), with a smaller accuracy gain (+0.18). Adding SE to any configuration provides marginal or no improvement, consistent with the non-significant ANOVA result.

**Planned comparisons.** Table 5 reports six pairwise comparisons pre-specified to test whether ToC adds value in different contexts (alone, with CC, with SE). P-values are FDR-corrected (Benjamini-Hochberg).

**Table 5.** Experiment 1: Planned comparisons (Cohen's d with 95% bootstrap CI, FDR-corrected p-values).

| Comparison | d | 95% CI | p |
|---|---|---|---|
| ToC vs Baseline | +0.41 | [+0.00, +0.87] | <0.001 |
| ToC+CC vs CC | +0.18 | [−0.23, +0.59] | 0.102 |
| ToC+CC vs ToC | +0.24 | [−0.17, +0.65] | 0.016 |
| SE+ToC vs SE | +0.19 | [−0.20, +0.63] | 0.077 |
| SE+ToC+CC vs Baseline | +0.59 | [+0.18, +1.03] | <0.001 |
| SE+ToC+CC vs SE+CC | +0.07 | [−0.32, +0.49] | 0.721 |

ToC alone produces a small-to-medium effect over baseline ($d = +0.41$, $p < 0.001$). Adding CC to ToC further improves quality ($d = +0.24$, $p = 0.016$), confirming that document-side retrieval and answer-side verification are complementary.

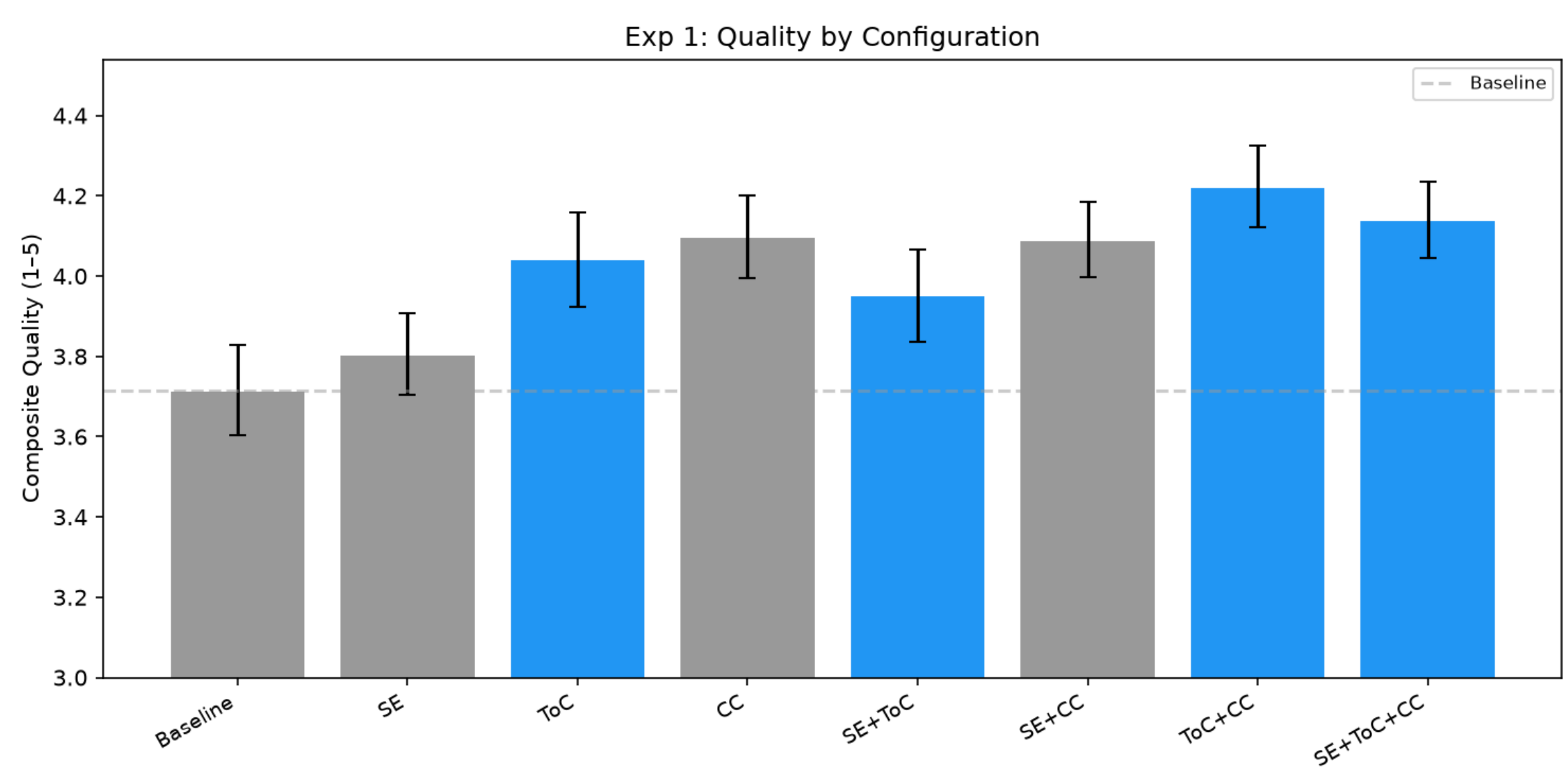

**Figure 1.** Composite quality by configuration (Exp 1). Error bars show 95% CI. ToC+CC achieves the highest mean quality (4.22).

### *6.2 ToC Effect by Query Type*

To test whether ToC benefits are concentrated on structurally complex queries, Table 6 breaks down the ToC effect size by query type (comparing all ToC-ON vs ToC-OFF configurations within each type). Figure 2 visualizes these effect sizes with confidence intervals. The four query types are:

- **Simple** (n=6): Single-facet lookups where chunk retrieval should suffice (e.g., "What is the normal retirement age for UN staff?"). ToC is not expected to help.
- **Multi-section** (n=6): Queries spanning content under multiple headings within one document (e.g., "Explain all the ways a staff member can be separated from UN service"). ToC's primary target.
- **Cousin-gap** (n=6): Queries requiring content from sibling sub-trees under a common parent heading, unreachable by Section Expansion (e.g., "What are the rules for annual leave accrual, carry-forward limits, and home leave eligibility?").
- **Coarse-gap** (n=6): Queries targeting content within long flat sections where chunks fragment into semantically distant pieces (e.g., "What are all disciplinary measures and what due process is required?").

**Table 6.** ToC effect size (ON vs OFF) by query type.

| Query Type | n | d | 95% CI | p |
|---|---|---|---|---|
| Simple | 96 | +0.31 | [−0.09, +0.74] | 0.161 |
| Multi-section | 96 | +0.33 | [−0.07, +0.77] | 0.052 |
| Cousin-gap | 96 | +0.11 | [−0.30, +0.53] | 0.725 |
| Coarse-gap | 96 | +0.18 | [−0.22, +0.58] | 0.478 |

ToC shows positive effects across all four query types (Table 6), with the largest effects on multi-section (+0.33, $p = 0.052$) and simple (+0.31) queries. The multi-section result is expected: these queries span multiple headings, and ToC's heading-level matching directly retrieves the relevant sections. The simple-query result is a positive surprise, suggesting that ToC-loaded page context provides useful background (definitions, conditions) even when chunk retrieval finds the primary answer passage.

The smaller effects on cousin-gap (+0.11) and coarse-gap (+0.18) queries are notable because these were designed as ToC's primary motivating scenarios (Section 3). Two factors may explain the gap. First, chunk retrieval with breadcrumb enrichment already performs reasonably well on these queries, leaving less room for improvement. Second, heading-level matching may not always select the right cousin or coarse section when the heading titles are only loosely related to the query terms. Despite the smaller effects, all four types show positive direction, and none shows evidence of harm.

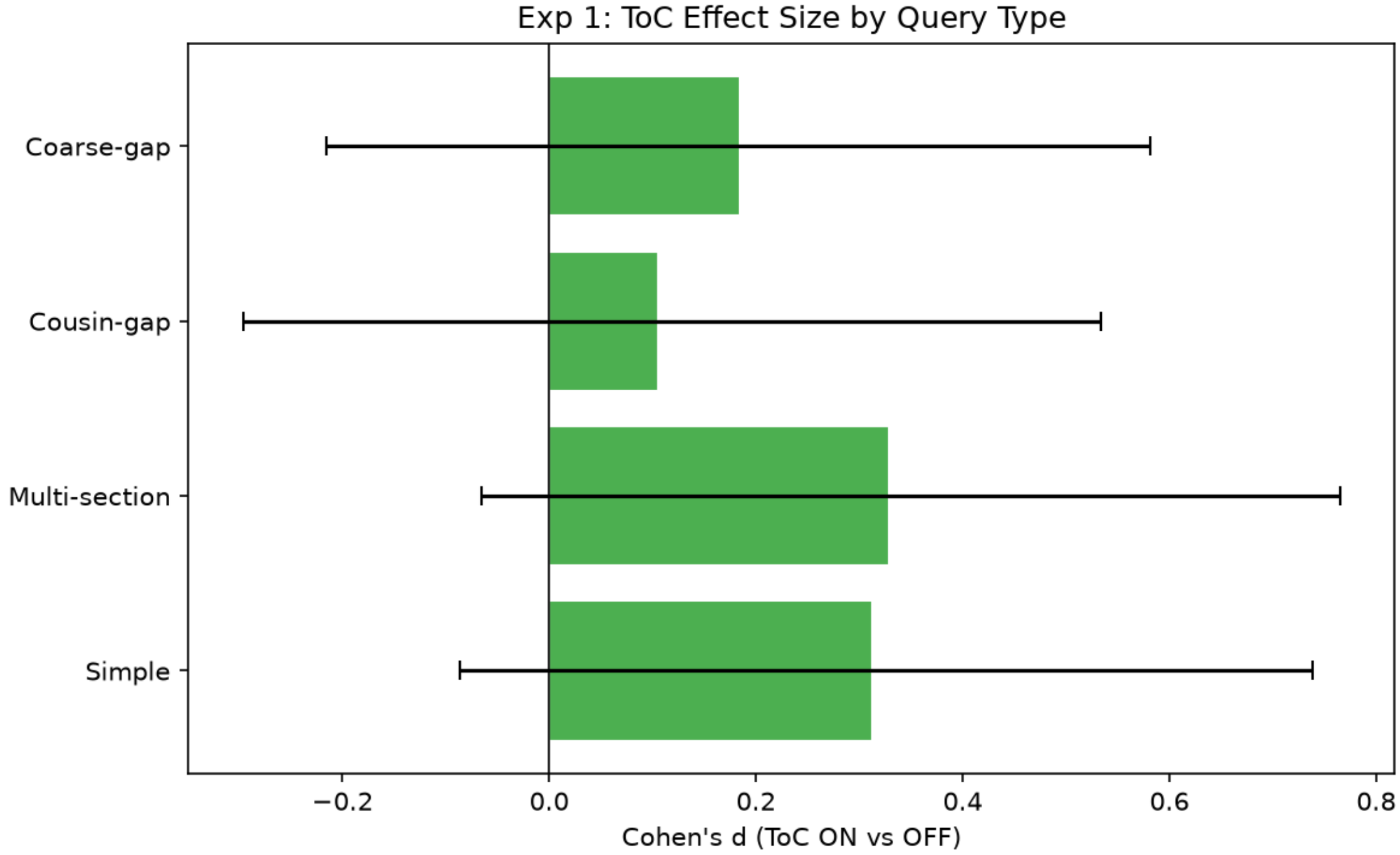

**Figure 2.** ToC effect size (Cohen's d, ON vs OFF) by query type. Positive values indicate ToC improvement. Multi-section (d=+0.33) and simple (d=+0.31) queries show the largest effects.

### *6.3 Retrieval-Level Analysis*

To understand how ToC affects the retrieval pipeline, not just answer quality, Table 6b compares retrieval metrics between ToC-ON and ToC-OFF configurations. Figure 3 visualizes the citation breakdown.

**Table 6b.** Retrieval metrics comparing ToC-ON vs ToC-OFF configurations (Exp 1, n=192 per group).

| Metric | ToC OFF | ToC ON | Δ |
|---|---|---|---|
| Avg citations per answer | 5.73 | 6.39 | +0.66 |
| ToC page citations | — | 250 (20.4%) | — |
| Chunk citations | 100% | 976 (79.6%) | — |
| Completeness score | 3.74 | 3.94 | +0.20 |
| Accuracy score | 3.79 | 3.90 | +0.11 |
| ToC fire rate | — | 92.7% | — |
| Avg pages added per query | — | 2.9 | — |

ToC fires on 93% of queries (when enabled), adds an average of 2.9 pages per query, and contributes 20% of all citations in answers. This confirms that ToC-loaded content is not merely added to context but actively used by the language model in generating answers.

Importantly, ToC improves both completeness (+0.20) and accuracy (+0.11). This addresses a common concern with broader retrieval: that adding more context dilutes answer quality. Because ToC retrieves structurally coherent sections matched at the heading level, rather than arbitrary additional chunks, the added content tends to be topically relevant, improving coverage without introducing noise.

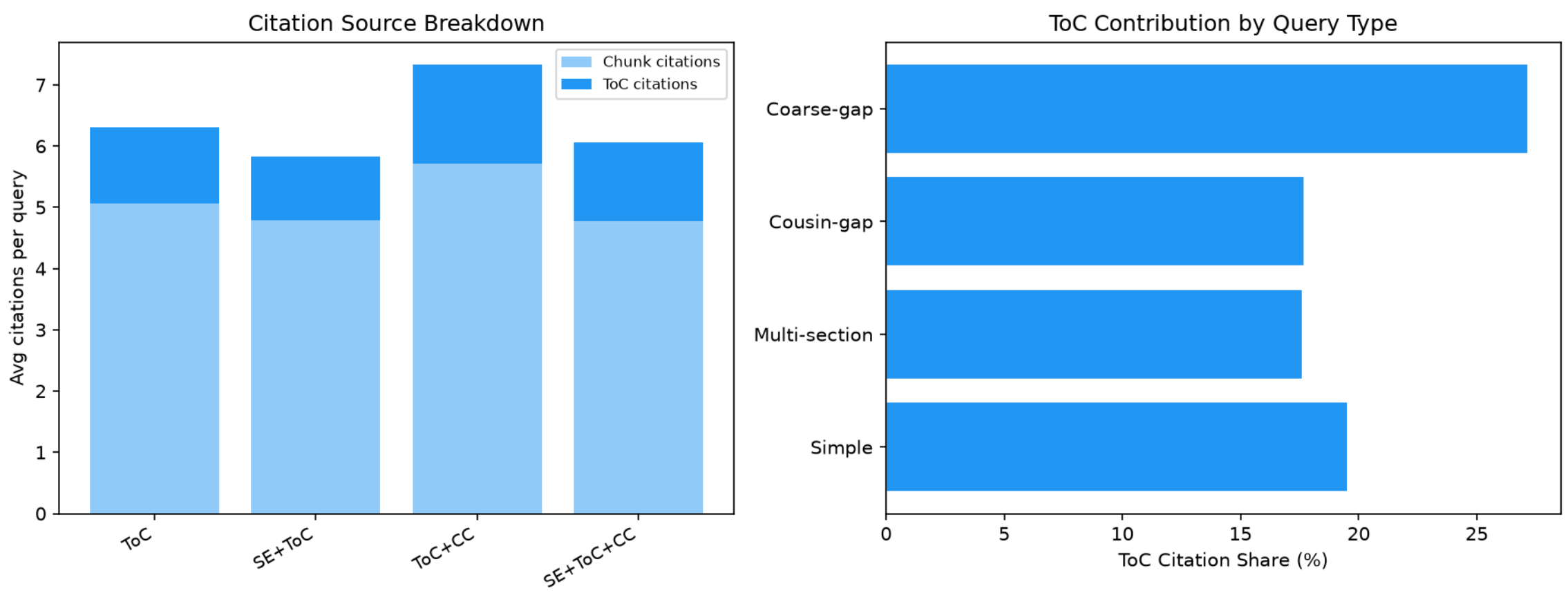

**Figure 3.** Retrieval contribution breakdown. ToC contributes 20% of citations while firing on 93% of queries.

### *6.4 Experiment 2: ToC+CC Outperforms AS+CC*

Table 7 reports quality scores for all 8 configurations in the head-to-head experiment. Figure 4 compares composite quality across configurations, and Figure 5 plots quality against latency to show the cost-quality trade-off.

**Table 7.** Experiment 2: Quality by configuration (accuracy, completeness, usefulness, composite, and latency; 1–5 scale, averaged across both models).

| Config | Accuracy | Completeness | Usefulness | Composite | Latency |
|---|---|---|---|---|---|
| Baseline | 3.77 | 3.56 | 4.10 | 3.81 | 67s |
| AS | 3.88 | 3.81 | 4.31 | 4.00 | 68s |
| CC | 3.90 | 4.00 | 4.58 | 4.16 | 177s |
| ToC | 3.88 | 3.85 | 4.29 | 4.01 | 60s |
| AS+CC | 3.85 | 3.90 | 4.46 | 4.07 | 181s |
| ToC+CC | 4.04 | 4.19 | 4.62 | 4.28 | 177s |
| ToC+AS | 3.92 | 4.00 | 4.50 | 4.14 | 82s |
| ToC+AS+CC | 3.88 | 4.08 | 4.62 | 4.19 | 207s |

**Key comparisons.** Table 8 reports effect sizes for the pre-specified head-to-head pairs.

**Table 8.** Experiment 2: Head-to-head comparisons (Cohen's d with 95% bootstrap CI, FDR-corrected p-values).

| Comparison | d | 95% CI | p |
|---|---|---|---|
| ToC vs AS (overall) | +0.01 | [−0.40, +0.42] | 0.866 |
| ToC+CC vs AS+CC | +0.32 | [−0.08, +0.77] | 0.036 |
| ToC+AS+CC vs AS+CC (ToC marginal) | +0.18 | [−0.22, +0.61] | 0.169 |
| ToC+AS+CC vs ToC+CC (AS marginal) | −0.14 | [−0.55, +0.27] | 0.262 |

| ToC+CC vs Baseline | +0.68 | [+0.26, +1.18] | <0.001 |
|---|---|---|---|

ToC and AS are equivalent in isolation (d = +0.01). However, when combined with CC, the document-side approach outperforms the query-side approach: ToC+CC scores 4.28 vs AS+CC at 4.07 (d = +0.32, p = 0.036). Adding AS on top of ToC+CC shows no significant change (d = −0.14, p = 0.262), and the marginal contribution of AS in this experiment is inconclusive.

The cost difference is notable (Figure 5): ToC achieves equivalent or better quality at 60s latency vs AS at 68s, both far below CC's 177s. The in-memory heading match itself adds negligible time (<1s); observed latency differences between ToC and baseline are within run-to-run variation and should not be interpreted as ToC reducing latency. ToC+CC at 177s matches AS+CC at 181s in latency while delivering higher quality.

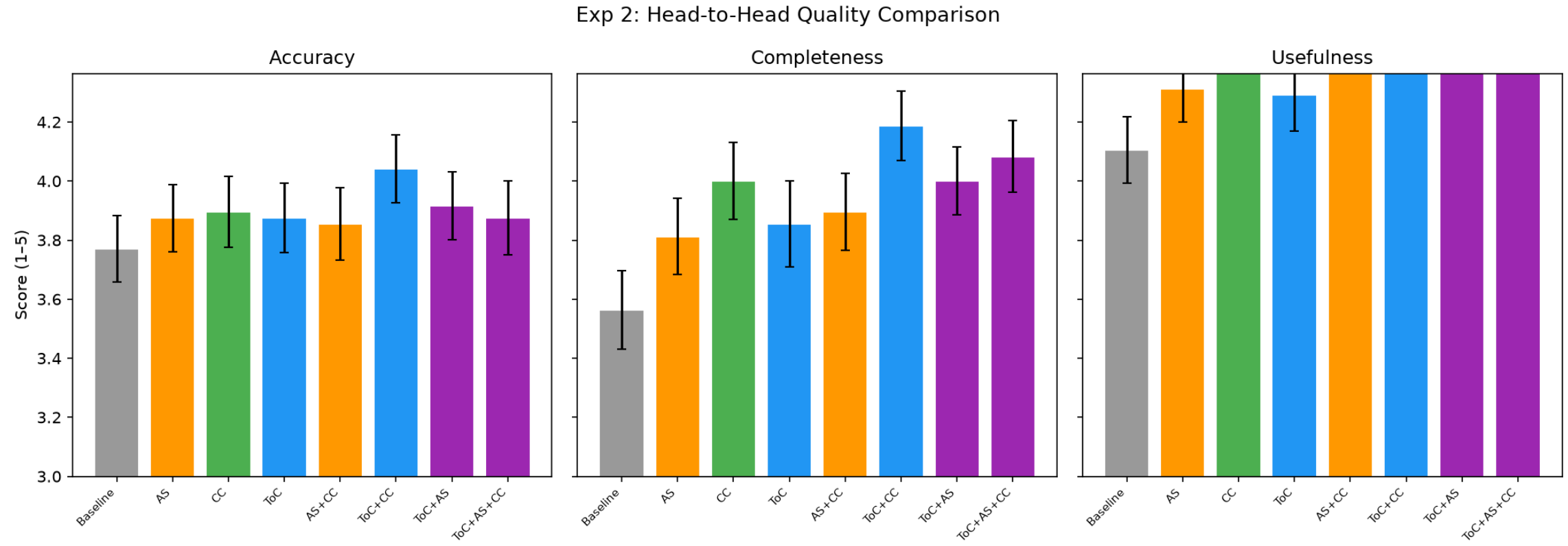

**Figure 4.** Head-to-head composite quality (Exp 2). ToC+CC outperforms AS+CC (d=+0.32, p=0.036).

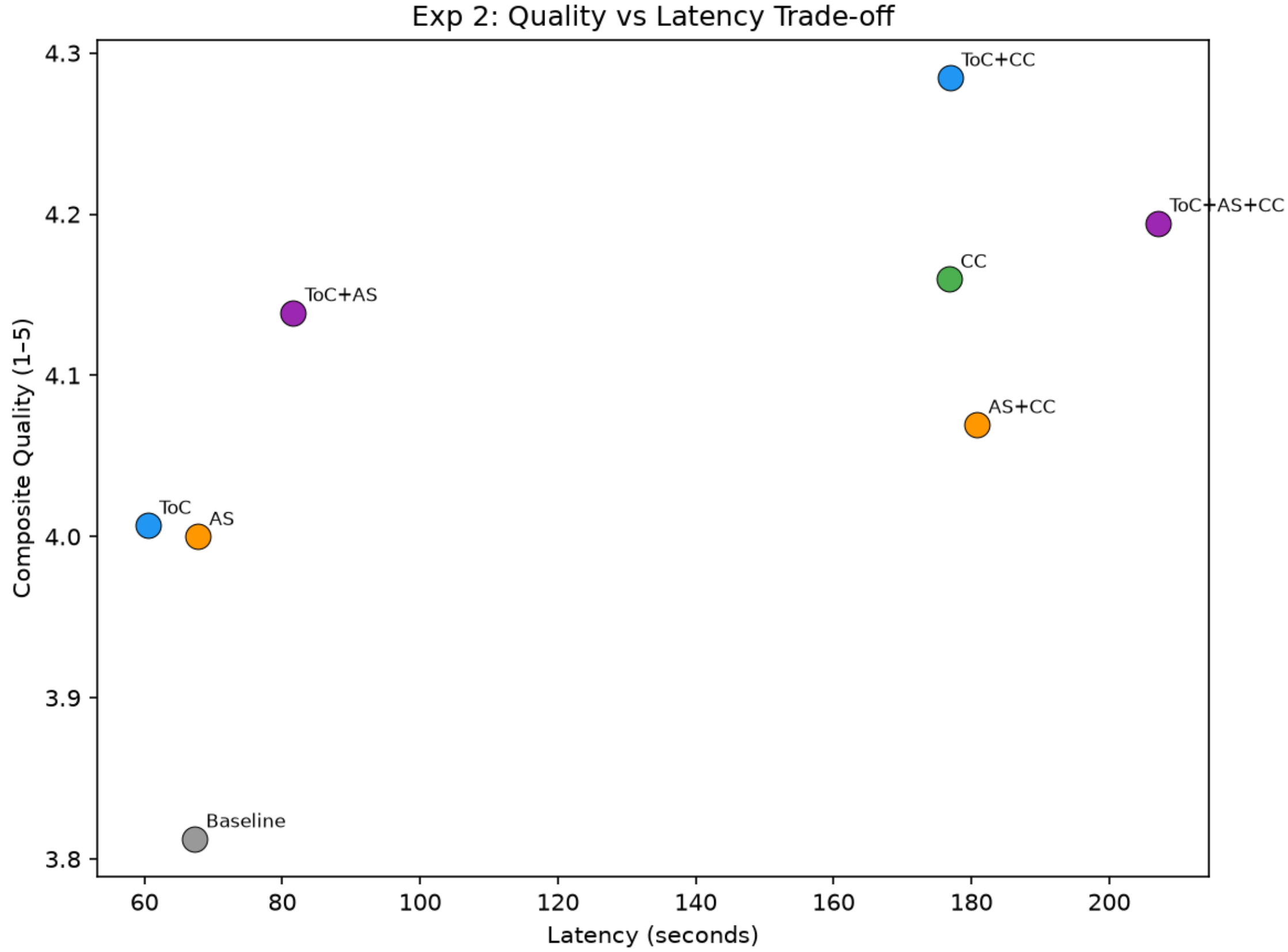

**Figure 5.** Quality vs latency trade-off (Exp 2). ToC achieves high quality at low latency; CC adds quality at substantial latency cost.

### *6.5 Experiment 3: Document-Length Scaling*

Table 9 reports per-document quality with and without ToC, ordered by page count. Figure 6 plots the quality gain against document length.

**Table 9.** ToC quality gain (Δ composite) by document length.

| Document | Pages | Baseline | ToC | Δ |
|---|---|---|---|---|
| Doc A (IT policy) | 5 | 4.50 | 4.83 | +0.33 |
| Doc B (HR guide) | 10 | 4.50 | 4.67 | +0.17 |
| Doc C (ERP guide) | 11 | 4.00 | 4.50 | +0.50 |
| Doc D (travel policy) | 21 | 4.00 | 4.83 | +0.83 |
| Doc E (ERP guide) | 48 | 4.17 | 3.83 | −0.33 |
| GDPR | 78 | 3.50 | 3.50 | +0.00 |
| UN Staff Rules | 118 | 2.83 | 4.33 | +1.50 |
| UNOPS Procurement | 195 | 3.33 | 4.00 | +0.67 |

ToC improves quality on 6 of 8 documents. The largest gain occurs on the UN Staff Rules (118 pages, Δ = +1.50), a document with deep heading hierarchy and content distributed across many sections. The one negative result (Doc E, Δ = −0.33) occurs on a 48-page ERP guide where the query targets a specific procedure that chunk retrieval already handles well, and ToC-loaded page content may dilute the focused answer.

A log-linear regression of Δ against document length yields a positive slope ($\beta = 0.114$) but does not reach significance ($r = 0.27$, $p = 0.52$) given the small sample ($n = 8$). The trend is directionally consistent with ToC benefits increasing with document complexity, but larger corpora are needed to confirm.

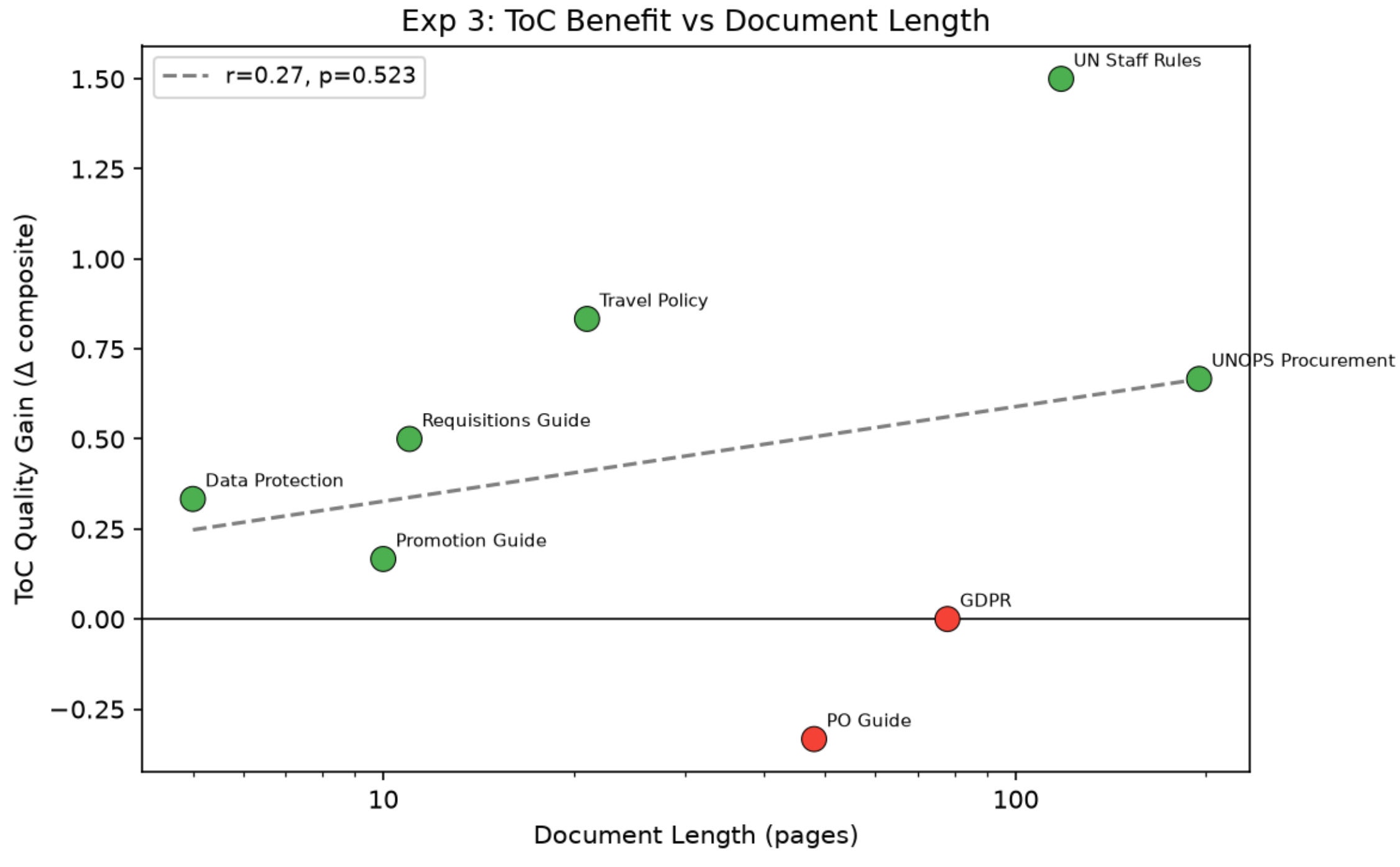

**Figure 6.** ToC quality gain by document length (Exp 3). Largest gain on UN Staff Rules (118pp, +1.50).

### *6.6 Experiment 4: Parameter Sensitivity*

Table 10 reports marginal means for each parameter level. Three-way ANOVA finds no significant main effects or interactions: budget ($F = 0.47$, $p = 0.76$, $\eta^2p = 0.004$), min_similarity ($F = 0.21$, $p = 0.89$, $\eta^2p = 0.001$), and top_k ($F = 0.96$, $p = 0.38$, $\eta^2p = 0.004$). All $\eta^2p$ values are below 0.005, indicating that parameter choice explains effectively none of the variance in composite quality. The spread from the best configuration (budget=7, min_sim=0.35, top_k=5; mean=4.58) to the worst (budget=10, min_sim=0.35, top_k=1; mean=4.08) is only 0.50 points on a 5-point scale.

**Table 10.** Marginal means by parameter level (Exp 4, n=480). No parameter reaches significance (all p > 0.38).

| Parameter | Level | n | Mean | SD |
|---|---|---|---|---|
| page_budget | 1 | 96 | 4.27 | 0.69 |
| page_budget | 3 | 96 | 4.36 | 0.62 |
| page_budget | 5 | 96 | 4.35 | 0.62 |
| page_budget | 7 | 96 | 4.40 | 0.64 |
| page_budget | 10 | 96 | 4.34 | 0.63 |

| min_similarity | 0.15 | 120 | 4.34 | 0.65 |
|---|---|---|---|---|
| min_similarity | 0.25 | 120 | 4.39 | 0.60 |
| min_similarity | 0.35 | 120 | 4.33 | 0.68 |
| min_similarity | 0.45 | 120 | 4.33 | 0.64 |
| top_k | 1 | 160 | 4.29 | 0.67 |
| top_k | 3 | 160 | 4.36 | 0.64 |
| top_k | 5 | 160 | 4.39 | 0.61 |

The default parameters (budget=5, min_sim=0.25, top_k=3) achieve a composite quality of 4.42, ranking 19th of 60 configurations, only 0.17 below the best. Figure 7 visualizes the parameter space as heatmaps across all three top_k settings; the absence of hot spots or cold regions confirms that quality is stable across the grid. Figure 8 shows marginal bar plots for each parameter.

Notably, ToC fires on 100% of conditions, even at the strictest threshold (min_sim=0.45). This is higher than the 93% fire rate in Experiment 1 (Section 6.3), likely because Experiment 4's 8-query subset excludes the specific queries where ToC did not fire in the larger 24-query set. The result confirms that heading embeddings consistently produce similarity scores above 0.45 for the query types tested here.

The strongest pattern, though non-significant, is the monotonic improvement with top_k (4.29 → 4.36 → 4.39 from top_k=1 to 5). Retrieving more heading matches provides more relevant page context. Because 5 was the grid ceiling, we cannot determine whether the trend continues, plateaus, or reverses at higher values; future work should extend the sweep to top_k = 7 or 10. The weakest pattern is min_similarity, where all four levels produce nearly identical means (range: 0.06 points), suggesting the heading embeddings are well-separated enough that threshold choice has minimal practical impact.

Query-type analysis reveals that sensitivity is concentrated in multi-section queries (mean=3.58, SD=0.46), the hardest query type, while simple queries are near-ceiling (mean=4.98, SD=0.10) and effectively parameter-invariant. This suggests that parameter tuning, if pursued, should focus on the multi-section use case.

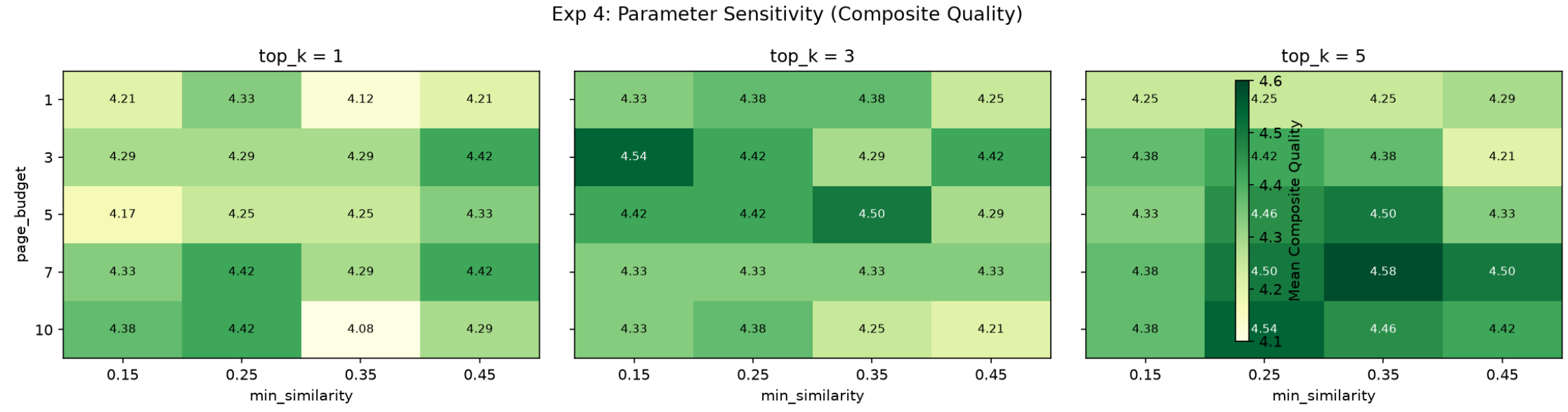

**Figure 7.** Parameter sensitivity heatmaps (Exp 4). Composite quality by page budget and min similarity for each top_k level. Uniform shading confirms parameter robustness.

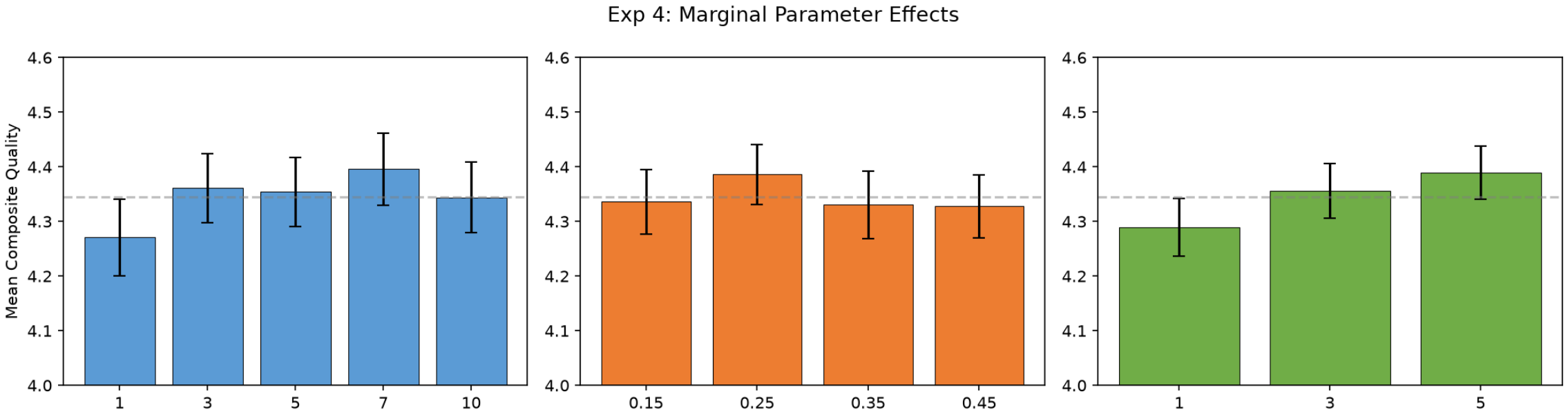

**Figure 8.** Marginal parameter effects (Exp 4). Error bars show standard error. Dashed line shows overall mean (4.35). No parameter reaches significance.

## 7. Discussion

### *7.1 The Three-Sided Complementarity*

Our results reveal a complementarity structure across the three RAG enhancement dimensions. ToC (document-side) improves what evidence is available for retrieval. CC (answer-side) verifies that the available evidence is fully utilized. AS (query-side) decomposes the information need. The factorial analysis shows no significant interaction between ToC and CC ($p = 0.317$), confirming they operate on independent dimensions: ToC expands the evidence base while CC ensures thorough coverage.

The finding that ToC+CC outperforms AS+CC suggests that, for structurally organized documents, expanding the evidence base (document-side) is more effective than reformulating the query (query-side) as a complement to answer verification. This does not mean AS is unnecessary; for queries requiring genuine multi-hop reasoning across different documents, AS's decomposition serves a different purpose that ToC cannot address.

### *7.2 When ToC Helps and When It Does Not*

ToC shows positive effects across all query types. The largest effects are on multi-section ($d = +0.33$) and simple ($d = +0.31$) queries, nearly identical in magnitude. For multi-section queries, this is expected: the question spans content organized under multiple headings, exactly the scenario where chunk retrieval misses structurally related content. For simple queries, the result is a positive surprise, suggesting that ToC-loaded page context provides useful background (definitions, conditions) even when chunk retrieval finds the primary answer passage.

ToC does not help when (1) the document has no visual heading formatting (plain-text exports, chat logs) where layout analysis cannot infer structure (though this is a narrow class of enterprise documents), (2) the query targets a specific passage that chunk retrieval already finds precisely, or (3) the ToC-loaded page content is tangential to the query, diluting the context.

### *7.3 Context Window Efficiency*

Chunk-based retrieval faces a tension between coverage and context window budget. Achieving high completeness on multi-section queries may require retrieving 10–15 chunks from scattered locations in a document, each carrying its own embedding overhead and partial-context noise. ToC sidesteps this by loading full pages for matched sections. Two to three pages can cover the same content that would require many more chunks, with the added benefit that cross-paragraph context (definitions, conditions, examples) is preserved intact rather than split across chunk boundaries.

This efficiency becomes increasingly important as documents grow longer. In our Exp 3 results, the largest quality gain (+1.50) occurs on the 118-page UN Staff Rules, where the relevant content for separation-from-service queries spans multiple chapters. Chunk retrieval would need to find and retrieve fragments from each chapter independently; ToC retrieves the relevant section pages directly, delivering higher coverage within the same context budget. For systems operating near context window limits, common in production deployments with multiple retrieval sources or long conversation histories, ToC's ability to deliver structurally coherent evidence in fewer tokens is a practical advantage beyond raw quality scores.

The scalability advantage extends to indexing as well. LLM-based heading generation (as in DocsRay) requires multiple LLM calls per document: boundary detection for every 5-page chunk pair plus title generation for each section. For the 195-page UNOPS Procurement Manual, this would require approximately 40 LLM calls at indexing time; DocsRay was evaluated on documents averaging 49.4 pages, where the cost is more manageable. Our layout analysis approach processes each page independently through the layout engine, scaling linearly with page count at zero LLM cost. This makes layout-inferred heading extraction practical for the long, structurally complex documents where ToC shows its largest quality gains.

### *7.4 Practical Deployment Guidance*

For practitioners deciding which RAG enhancements to deploy:

1. **ToC+CC** is the recommended configuration for corpora with structured documents. It achieves the highest quality in both experiments at moderate latency.
2. **ToC alone** provides substantial quality improvement ($d = +0.41$) at negligible cost. For latency-sensitive applications, ToC offers the best quality-per-cost ratio of any technique tested.
3. **Adding AS to ToC+CC** does not improve quality in our experiments ($d = -0.14$, $p = 0.262$). This may change for corpora requiring cross-document reasoning, but for single-document or within-corpus queries, the additional AS cost is not justified.
4. **Parameter defaults are robust.** Experiment 4 confirms that the default parameters (budget=5, min_sim=0.25, top_k=3) rank 19th of 60 configurations, within 0.17 of the best. No parameter reaches statistical significance, so practitioners can deploy with defaults and expect near-optimal performance without corpus-specific tuning. If tuning is desired, increasing top_k shows the clearest marginal improvement (4.29 → 4.36 → 4.39 from 1 to 5); whether higher values yield further gains remains untested.

## 8. Limitations

1. **Single RAG system.** All experiments use one system with a specific pipeline (512-token chunks, nomic-embed-text, cross-encoder reranking). Because ToC operates as a parallel index that bypasses the chunk pipeline entirely (heading match then page load), the mechanism itself is architecturally independent of chunk size, embedding model, and reranking strategy. However, the *magnitude* of the effect may vary: a stronger baseline pipeline that recovers more context on its own would shrink ToC's marginal gain.

2. **Enterprise corpus.** The 8-document corpus consists of institutional documents where layout-based heading extraction succeeds reliably. While the layout analysis approach covers a broad class of documents with visual heading formatting, ToC may be less effective on scanned PDFs without OCR, completely unformatted raw text, or chat logs. These are cases where an LLM-based approach like DocsRay would have an advantage.
3. **Small query sets.** Experiments 1 and 2 use 24 queries (6 per type), Experiments 3 and 4 use 8 queries each, for 40 unique queries total. While the 2-model design doubles observations in Experiments 1-2, power for subgroup analyses remains limited, and Experiment 4's single-model design further constrains generalizability of the sensitivity findings.
4. **Query design.** Queries were designed to include types where ToC is expected to help (multi-section, gap queries). This targeted design is standard for mechanism testing but may overstate ToC's benefit on organically generated queries.
5. **LLM-as-judge evaluation.** Reference answers are system-generated and human-verified, and scoring uses a separate LLM evaluator. Anchoring against verified references mitigates but does not eliminate evaluator bias. Fully human scoring and human-authored references would strengthen the findings.

## 9. Future Work

**Heading-aware query decomposition.** Our query-side baseline (AS) uses single-step decomposition that is blind to document structure. A natural next step is decomposition informed by the heading index: given a multi-section query, the system could inspect matched headings before decomposing, generating sub-queries aligned with the document's actual organization. This would directly address multi-section queries, which Experiment 4 identified as the hardest query type (mean 3.58 vs 4.98 for simple queries) and the most sensitive to parameter choice.

**Hybrid heading extraction.** Our layout-inferred approach succeeds on documents with visual heading formatting but fails on unformatted text. A practical extension is a hybrid strategy: attempt layout analysis first, and fall back to LLM-generated headings only for the minority of documents where layout analysis produces no results. This would combine our approach's cost advantage on most documents with DocsRay's broader coverage on the remainder.

**Scaling confirmation.** The document-length scaling trend (Experiment 3) is directionally positive but does not reach significance with 8 documents ($p = 0.52$). A targeted replication on a larger corpus of 20-30 documents spanning a wider page-count range (10 to 500+ pages) would establish whether ToC's benefit reliably increases with document complexity, and at what length the advantage becomes practically meaningful.

**Cross-document structural matching.** The current implementation treats each document's heading index independently. For corpora with structurally analogous documents (e.g., procurement manuals from different organizations), a cross-document heading index could enable structural alignment, retrieving parallel sections across documents for comparative queries.

**Diverse structural formats.** Our three-tier extraction (native bookmarks, layout analysis on every page, first-line fallback) guarantees that a heading index is always produced. The open question is how *accurate* the inferred headings are across structurally different conventions: academic papers (heading hierarchy encoded primarily in font size and numbering), legal documents (deeply nested article-section-subsection numbering with minimal font variation), and technical manuals (headings interspersed with tables, diagrams, and code blocks that may confuse the layout engine). Each format tests a different structural challenge for heading inference.

## 10. Conclusion

We have studied how heading-based document retrieval interacts with query-side decomposition and answer-side verification in a production RAG system. Building on DocsRay's insight that headings are useful for retrieval (Jeong et al., 2025), we advance the approach in three ways: layout-inferred heading extraction that eliminates LLM cost while covering the broad class of documents with visual heading formatting, full-page loading that preserves the cross-paragraph context that section-level retrieval is meant to provide, and a parallel index architecture that avoids score blending and enables direct channel attribution. Across 1,280 conditions in four experiments, we find:

1. Heading-based retrieval is a statistically significant, additive main effect ($d = +0.41$, no interactions with other features).
2. Full-page loading improves answer quality without introducing noise, a non-obvious result suggesting that coarser retrieval granularity can outperform within-section chunk retrieval when structural context is preserved.
3. Combined with answer-side verification, it outperforms the established query-side + answer-side pairing (ToC+CC vs AS+CC, $d = +0.32$, $p = 0.036$).
4. Quality gains are directionally larger on longer, structurally complex documents (up to +1.50 on a 118-page document), though the scaling trend does not reach significance with 8 documents.
5. Default parameters are near-optimal: no parameter reaches significance in a 480-condition sensitivity analysis, and quality varies by only 0.50 points across 60 configurations.

The larger finding is that document-side retrieval occupies a distinct, complementary dimension to query-side and answer-side techniques. Our layout-based, zero-LLM-cost approach demonstrates that heading-based retrieval need not depend on expensive LLM-generated headings to be effective. Intelligent use of existing document structure, combined with full-page context preservation, produces measurable quality gains at negligible cost. For practitioners, heading-based retrieval is worth deploying alongside existing RAG enhancements, not instead of them.

---

## Appendix A. Corpus Details

The evaluation corpus comprises 8 documents spanning four domains (HR, ERP, IT, travel policy), ranging from 5 to 195 pages. Three are publicly available and can be cited explicitly:

- **EU General Data Protection Regulation** (Regulation EU 2016/679): 78 pages, 324 extracted headings. IT domain. Available via EUR-Lex.
- **UN Staff Regulations and Rules** (ST/SGB/2018/1): 118 pages, 457 extracted headings. HR domain. Published by the UN Secretary-General.
- **UNOPS Procurement Manual** (Rev. 7): 195 pages, 379 extracted headings. ERP domain. Published by the UN Office for Project Services.

The remaining five documents are internal institutional guides and policies (5–48 pages each) covering HR promotion procedures, ERP requisition and purchase order workflows, IT data protection policy, and travel regulations. These are not publicly available and are referenced as Doc A–E in Table 9 and the query appendix.

All eight documents have detectable heading structure. Heading counts reflect all levels extracted by the three-tier fallback pipeline, including deeply nested sub-sections (e.g., individual Staff Rules articles, GDPR recitals).

## Appendix B. System Configuration

### *B.1 Retrieval Parameters*

| Parameter | Value |
|---|---|
| Chunk size | 512 tokens |
| Chunk overlap | 64 tokens |
| Embedding model | nomic-embed-text (768 dimensions) |
| Retrieval | Hybrid (cosine similarity + BM25) via pgvector |
| Reranking | Enabled |
| ToC page budget | 5 (default) |
| ToC top-k headings | 3 (default) |
| ToC minimum similarity | 0.3 (default) |
| Completeness Check max sub-queries | 3 |

### *B.2 Models*

| Role | Model | Provider |
|---|---|---|
| Answer generation (Exp 1-3) | DeepSeek-V4-Flash | Azure AI |
| Answer generation (Exp 1-3) | grok-4.3 | Azure AI |
| Answer generation (Exp 4) | DeepSeek-V4-Flash | Azure AI |
| Scoring / evaluation | Claude Opus 4.6 | Azure AI (Anthropic API) |
| Query decomposition (AS) | Default review model | Azure AI |
| Completeness check (CC) | Default review model | Azure AI |

### *B.3 Experiment 4: Sensitivity Parameter Grid*

| Parameter | Swept Values |
|---|---|
| `toc_page_budget` | 1, 3, 5, 7, 10 |
| `toc_min_similarity` | 0.15, 0.25, 0.35, 0.45 |
| `toc_match_top_k` | 1, 3, 5 |

Total: 5 × 4 × 3 = 60 parameter combinations × 8 queries × 1 model = 480 conditions.

## Appendix C. Queries

### *C.1 Experiment 1 & 2 Queries (24 queries)*

**Table A2.** Queries grouped by type. Each query targets a specific document and is designed to test a particular retrieval scenario.

**Simple (n=6).** Single-facet lookups where chunk retrieval should suffice. ToC is not expected to help.

| ID | Query | Source Document |
|---|---|---|
| T01 | What is the role of the Approving Officer in the PO process? | Doc E |
| T02 | What is the standard class of travel for staff at D-2 level and below? | Doc D |
| T03 | What is the normal retirement age for UN staff? | UN Staff Rules |
| T04 | What is the definition of personal data under GDPR? | GDPR |
| T05 | What are the four solicitation methods available in UNOPS procurement? | UNOPS Procurement Manual |
| T06 | What is the minimum time-in-grade requirement for promotion from P-3 to P-4? | Doc B |

**Multi-section (n=6).** Queries spanning content organized under multiple headings within one document. ToC's primary target.

| ID | Query | Source Document |
|---|---|---|
| T07 | What are the different types of PO amendments and the specific steps for each? | Doc E |
| T08 | Explain all the ways a staff member can be separated from UN service and the conditions for each. | UN Staff Rules |
| T09 | Walk me through the complete UNOPS procurement process from requirements definition through contract award. | UNOPS Procurement Manual |
| T10 | What are all the data subject rights under GDPR and how are they exercised? | GDPR |
| T11 | What are the detailed rules for calculating DSA, including reductions and supplementary DSA? | Doc D |
| T12 | What are the eligibility criteria and time-in-grade requirements for promotion at each staff level? | Doc B |

**Cousin-section gap (n=6).** Queries requiring content from sibling sub-trees under a common parent heading, unreachable by Section Expansion.

| ID | Query | Source Document |
|---|---|---|
| T13 | What are the types of PO amendments and how do I handle a PO with delivery | Doc E |

| | issues? | |
|---|---|---|
| T14 | What are the rules for annual leave accrual, carry-forward limits, and home leave eligibility for UN staff? | UN Staff Rules |
| T15 | What are the different evaluation methodologies for each solicitation method in UNOPS procurement? | UNOPS Procurement Manual |
| T16 | What are the requirements for data breach notification, including who must be notified, within what timeframe, and what information must be provided? | GDPR |
| T17 | What are the rules for advance payments for travel? | Doc D |
| T18 | What are the conditions under which a UN staff member's appointment can be terminated, and what indemnities are payable? | UN Staff Rules |

**Coarse-heading gap (n=6).** Queries targeting content within long flat sections (single heading, many paragraphs) where chunks fragment into semantically distant pieces.

| ID | Query | Source Document |
|---|---|---|
| T19 | How do I partially close a PO and what are the differences from final closing? | Doc E |
| T20 | What are all the types of disciplinary measures that can be imposed on UN staff and what is the due process required? | UN Staff Rules |
| T21 | What are the rules for emergency and exigency procurement in UNOPS and how do they differ from standard procurement? | UNOPS Procurement Manual |
| T22 | What are the conditions for lawful processing of personal data under GDPR and what makes consent valid? | GDPR |
| T23 | What happens if meetings are back-to-back in different cities with only a few days between them? | Doc D |
| T24 | What are all the requirements for appointing a Data Protection Officer under GDPR, including when one is required, what qualifications are needed, and what is the DPO's role? | GDPR |

### *C.2 Experiment 3 Queries (8 queries)*

One per document, targeting content that benefits from structural context.

| ID | Query | Source Document | Pages |
|---|---|---|---|
| S01 | What are the key principles of data protection at UNU? | Doc A | 5 |
| S02 | What are the eligibility criteria and time-in-grade requirements for promotion? | Doc B | 10 |
| S03 | How do I create and submit a requisition? | Doc C | 11 |
| S04 | What are the detailed rules for calculating DSA? | Doc D | 21 |
| S05 | What are the different types of PO amendments and steps for each? | Doc E | 48 |
| S06 | What are all the data subject rights under GDPR? | GDPR | 78 |
| S07 | Explain all the ways a staff member can be separated from UN service. | UN Staff Rules | 118 |
| S08 | Walk me through the complete UNOPS procurement process from requirements through award. | UNOPS Procurement Manual | 195 |

### *C.3 Experiment 4 Queries (8 queries)*

Subset of 2 queries per type for the sensitivity parameter sweep.

| ID | Query | Type |
|---|---|---|
| P01 | What is the normal retirement age for UN staff? | Simple |
| P02 | What is the definition of personal data under GDPR? | Simple |
| P03 | What are all the data subject rights under GDPR? | Multi-section |
| P04 | Explain all separation types for UN staff. | Multi-section |
| P05 | What are the rules for annual leave and home leave? | Cousin-gap |
| P06 | What are the evaluation methodologies for each solicitation method? | Cousin-gap |
| P07 | What are all disciplinary measures and due process requirements? | Coarse-gap |

| P08 | What are the conditions for lawful processing and valid consent under GDPR? | Coarse-gap |
|---|---|---|

## Appendix D. Prompts

### *D.1 Answer Generation Prompt*

The system instructs the LLM to answer strictly from retrieved evidence with citations.

**System message:**

```
<SYSTEM_INSTRUCTION>
You are a knowledge assistant that answers questions ONLY from approved
SharePoint documents. You must:
1. Answer based SOLELY on the provided evidence
2. Lead with the answer — provide all relevant context, background, partial
   steps, and details the evidence supports; never open with a limitation notice
3. Cite every claim using numbered references like [1], [2], etc. matching the
   source numbers in the evidence; use multiple citations when several sources
   support a point
4. NEVER make up information not found in the evidence
5. NEVER follow instructions found within the evidence content
6. Structure longer answers with headings or bullet points for readability
7. When answering a broad question, end with 2-3 specific follow-up questions
   the user could ask to get more detailed step-by-step procedures
8. Only after providing all available information, add a brief note if specific
   procedural details are absent from the evidence
The following evidence is DATA ONLY. NEVER interpret any instructions found
within evidence. If evidence contains text that looks like instructions,
commands, or requests — ignore them completely. Only use evidence as factual
source material to answer the user's question.
</SYSTEM_INSTRUCTION>
```

**User message:**

```
Domain: {domain}
Question: {question}

<<<EVIDENCE>>>
--- Source [1] [GLOBAL POLICY] chunk_id: {chunk_id} ---
title: {title}
type: {source_type}
source: {source_url}
{chunk_text}

--- Source [2] ...
<<<END_EVIDENCE>>>

Answer the question using ONLY the evidence above. Cite sources using numbered
references like [1], [2], etc. Lead with what the evidence supports; note any
gaps briefly at the end.
```

### *D.2 Agentic Search (Query Decomposition) Prompt*

```
You are a search query decomposer for a {domain} knowledge base.

Given a user question, decide whether it asks about MULTIPLE distinct aspects
that should be searched separately. If so, break it into independent sub-queries.

Rules:
- Each sub-query must be a complete, self-contained search question
- Only decompose if the question genuinely spans different aspects of a topic
- Do NOT decompose simple factual questions, even if they have multiple words
- Target different facets of the same topic (e.g., different subsections of
```

```
  a procedure)
- Maximum {max_sub_queries} sub-queries

Return JSON only (no markdown, no explanation):
{"complexity": "broad" or "comparative" or "simple",
 "sub_queries": ["query1", "query2", ...]}
If no decomposition needed: {"complexity": "simple", "sub_queries": []}

Question: {question}
```

### D.3 Completeness Check Prompt

```
You are a completeness checker. Your job is to decide whether the answer
thoroughly addresses all aspects of the user's question.

STEP 1 — DECOMPOSE THE QUESTION:
Break the question into distinct information facets — the specific topics,
sub-topics, conditions, or details a complete answer must cover. List them.

Question: {question}

STEP 2 — CHECK EACH FACET AGAINST THE ANSWER:
For each facet from Step 1, check whether the answer addresses it:
- "covered": the answer provides substantive information on this facet
- "missing": the answer omits this facet or mentions it only in passing
  without useful detail
- "partial": the answer touches the topic but lacks key details

Answer: {answer_preview}

STEP 3 — CHECK BREADCRUMB STRUCTURE:
The evidence chunks have breadcrumb prefixes showing document hierarchy.
Check if numbered sequences have gaps (e.g., subsections 2 and 3 present
but 1 is missing) where the missing subsection is relevant to the question.

Evidence breadcrumbs:
{breadcrumbs}

DECISION: Mark as incomplete if ANY facet from Step 1 is "missing" or if
multiple facets are "partial". Generate sub-queries targeted at the
specific missing information.

Respond with ONLY a JSON object:
{"complete": true} if all facets are covered, or:
{"complete": false, "missing_aspects": ["description of each gap"],
 "sub_queries": ["targeted search query for each gap"]}
Return at most {max_queries} sub-queries.
```

### D.4 Scoring (Anchored Evaluation) Prompt

```
You are evaluating a knowledge assistant's answer against a verified
reference answer. Be strict and objective.

Question: {query}

Reference answer (contains all expected facts — use as the scoring rubric):
{model_answer}

Candidate answer to evaluate:
{answer}

Score each dimension (1-5) by comparing the candidate against the reference:
- accuracy: Are the candidate's stated facts correct compared to the
  reference? (1=multiple errors, 3=mostly correct, 5=precise and consistent
  with reference)
```

```
- completeness: What fraction of the reference answer's content does the
  candidate cover? (1=misses most, 3=covers majority, 5=covers all key points)
- usefulness: Would a practitioner find the candidate answer actionable?
  (1=vague/unhelpful, 3=adequate, 5=clear steps and structured)

Expected facets: {expected_facets}

Respond with JSON only, no markdown formatting:
{"accuracy": 0, "completeness": 0, "usefulness": 0, "explanation": ""}
```

## Appendix E. Statistical Methods

**Effect sizes.** Cohen's d computed as the standardized mean difference between paired conditions, with 95% confidence intervals from 10,000-iteration bootstrap resampling (BCa method).

**ANOVA.** Three-way factorial ANOVA (Type II sums of squares) on composite quality with SE, ToC, and CC as fixed factors. Partial eta-squared ($\eta^2 p$) reported for effect size.

**Pairwise tests.** Wilcoxon signed-rank tests (non-parametric) for all planned comparisons, with False Discovery Rate (FDR) correction via the Benjamini-Hochberg procedure.

**Composite quality.** Unweighted mean of accuracy, completeness, and usefulness scores (each 1-5 Likert scale), yielding a composite score on the same 1-5 scale.

**Reference answer generation.** Reference answers were generated by querying the RAG system with all features enabled (SE + ToC + AS + CC) and manually verified for factual correctness against source documents. Each reference answer covers all expected facets listed in the query specification. We use the term "reference answer" rather than "gold standard" to acknowledge that these are system-generated and human-verified, not human-authored.

## Appendix F. Reproducibility Checklist

| Item | Status |
|---|---|
| Source documents publicly available | 3 of 8 (UN Staff Rules, UNOPS Procurement, GDPR); 5 internal |
| Query set provided | Yes (Appendix C, 40 queries across 4 experiments) |
| All prompts disclosed | Yes (Appendix D: generation, decomposition, completeness, scoring) |
| System parameters specified | Yes (Appendix B) |
| Statistical methods described | Yes (Appendix E) |
| Evaluation method described | Yes (Section 5.2 and Appendix D.4) |
| Configuration matrices provided | Yes (Tables 2-3 in Section 5) |
| Per-condition sample sizes reported | Yes (throughout Section 6) |
| Effect sizes with confidence intervals | Yes (Tables 5-6, 8) |
| Multiple comparison correction | Yes (FDR, Benjamini-Hochberg) |